\documentclass[preprint,prl,epsf,aps]{revtex4}
\newcommand{\nix}[1]{}
\usepackage{bm}

\input epsf

\begin{document}

\title{ Resonant inversion of the circular photogalvanic effect in $n$-doped quantum
wells}

\author{S.D.~Ganichev\dag\ddag,  V.V.~Bel'kov\ddag,  Petra~Schneider\dag,  E.L.~Ivchenko\ddag, S.A.~Tarasenko\ddag, W.~Wegscheider\dag, D.~Weiss\dag,
D.~Schuh\S, E.V.~Beregulin\ddag,
and W.~Prettl\dag, }

\address{ \dag~ Fakult\"{a}t
Physik, University of
Regensburg, 93040, Regensburg, Germany\\
\ddag~A.F.~Ioffe Physico-Technical Institute,
194021 St.~Petersburg, Russia\\
\S~Walter Schottky Institute, TU Munich,  D-85748 Garching, Germany}

\date{\today}
\draft

\begin{abstract}

We show that the sign of the circular photogalvanic effect  can be
changed by tuning the radiation frequency of circularly polarized
light. Here resonant inversion of the photogalvanic effect has
been observed for direct inter-subband transition in $n$-type GaAs
quantum well structures. This inversion of the photon helicity
driven current is a direct consequence of the lifting of the spin
degeneracy due to {\boldmath$k$}-linear terms in the Hamiltonian
in combination with energy and momentum conservation and optical
selection rules.

\end{abstract}

\maketitle

\newpage

\section{Introduction}

Effects caused by spin-orbit interaction in compound semiconductor
heterojunctions have been the subject of a growing number of
investigations recently~\cite{Wolf01p1488,spintronicbook02}. In
two-dimensional systems based on quantum wells (QWs)
the electron spin couples to the electron motion and results,
under optical orientation with circularly polarized light, in spin
photocurrents~\cite{PRL01}. The direction and magnitude of this
spin photocurrent depend on the degree of circular polarization of
the incident light~\cite{APL00}. This phenomenon belongs to the
class of photogalvanic effects~\cite{book} and represents here a
circular photogalvanic effect (CPGE).

It was shown in~\cite{PRL01} that the CPGE in zinc-blende
structure based QWs is caused by spin orientation of carriers in
systems where  the spin degeneracy of the band structure is lifted
by {\boldmath$k$}-linear terms in the
Hamiltonian~\cite{Bychkov84p78,Dyakonov86p110}. In this case
homogeneous irradiation of QWs with circularly polarized light
results in an asymmetric distribution of photo-excited carriers in
{\boldmath$k$}-space which leads to the  current. So far this
effect has been observed only for indirect intra-subband
transitions in $n$- and $p$-type QWs (Drude absorption) and for
direct inter-subband heavy-hole - light-hole transitions in
$p$-type QWs~\cite{PRL01,PhysicaE02}.

Here we report on the first observation of a resonant inversion of
the CPGE at direct transitions between size quantized subbands in
$n$-type QWs. This effect demonstrates in a very direct way the
spin splitting of subbands in {\boldmath $k$}-space  in zero
electric and magnetic field due to spin-orbit interaction. We show
that the sign of the spin driven circular photogalvanic  current
can be reversed by tuning the radiation frequency. This inversion
of the photon helicity driven current is a direct consequence of
{\boldmath $k$}-linear terms in the subband structure in
combination with conservation laws and optical selection rules.

\section{Experimental results}

Direct inter-subband transitions between the lowest ($e1$) and the
second ($e2$) conduction subband in $n$-type GaAs QWs have been
obtained by applying a line tunable pulsed transversely excited
atmospheric pressure (TEA)-CO$_2$ laser. The laser   yields strong
linearly polarized emission at wavelengths $\lambda$ between
9.2\,$\mu$m  and 10.8\,$\mu$m corresponding to photon energies
$\hbar \omega$ ranging from  135~meV to 114~meV. The quantum  well
widths were chosen to be $\approx$8~nm, so that the separation of
the subbands $e1$ and $e2$ matches the photon energy range of the
laser~\cite{Hilber97p85,Tsujino00p1560}. Molecular-beam-epitaxy
grown (001)- and (113)-oriented  $n$-type GaAs/AlGaAs QW samples
of 8.8~nm, 8.2~nm and 7.6~nm width with free-carrier densities
ranging between $2\cdot10^{11}$\,cm$^{-2}$ and
$1\cdot10^{12}$\,cm$^{-2}$ were investigated at room temperature.
The (113)-oriented samples have been grown on GaAs (113)A
substrates employing growth conditions under which Si dopants are
predominantly incorporated as donors~\cite{Sakamoto95p1444} as
confirmed by Hall measurements.

\begin{figure}
 \centerline{\epsfxsize 90mm \epsfbox{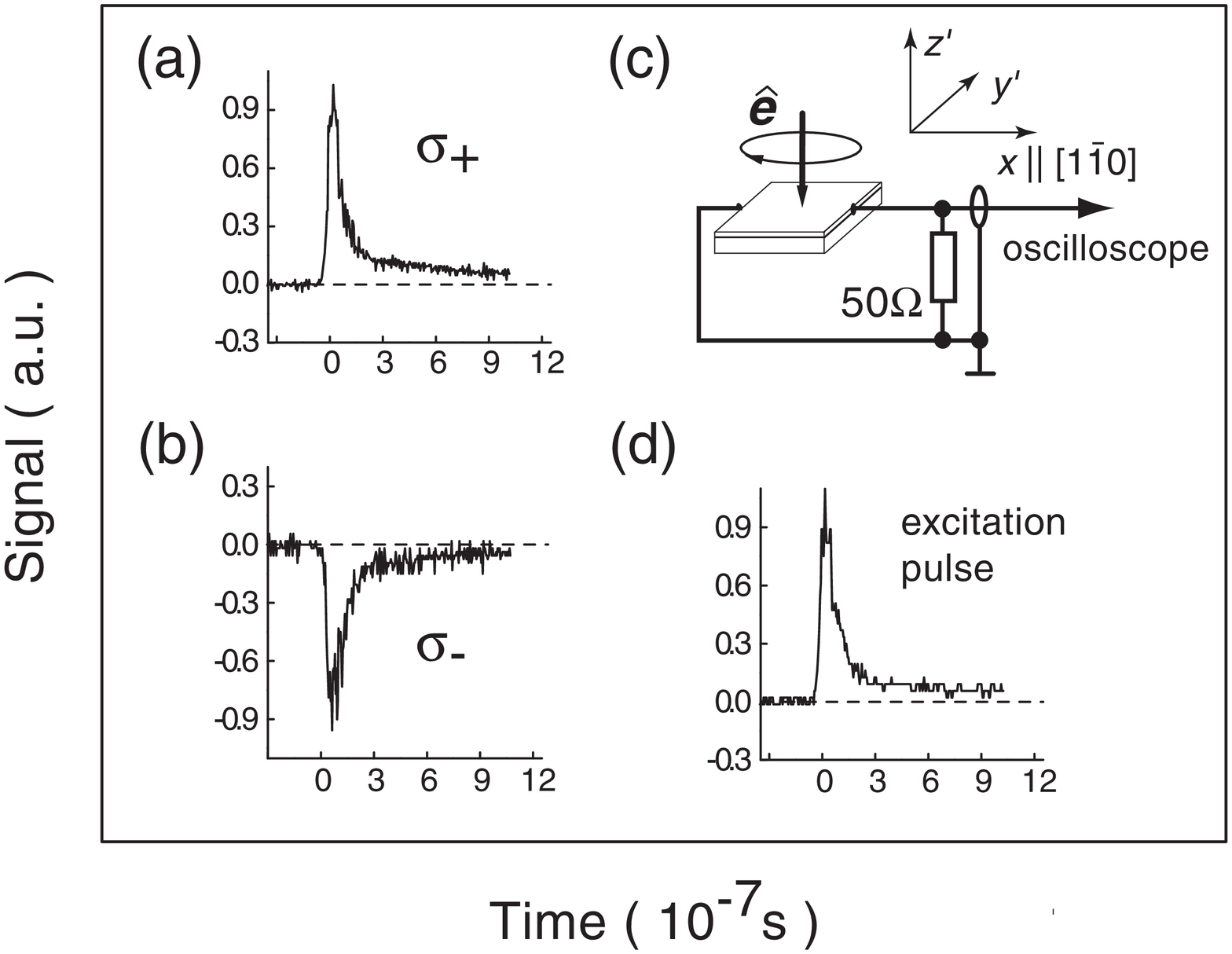  }}

\caption{Oscillographic traces obtained through excitation with
$\lambda = 10.6\,\mu$m radiation of (113)-grown $n$-GaAs QWs. (a)
and (b) show CPGE signals obtained for $\sigma_{+}$ and
$\sigma_{-}$-circular polarization, respectively. For comparison
in (d) a signal pulse of a fast photon drag detector is plotted.
In (c) the measurement arrangement is sketched. For (113)-grown
samples being of C$_s$ symmetry radiation was applied at normal
incidence and the current was detected in the direction $x
\parallel$ [1$\bar{1}$0]. For (001)-grown QWs oblique incidence
with light propagating along [110] direction was used and the
current was detected in $x \parallel$ [1$\bar{1}$0] direction.}
\label{fig1}
\end{figure}

On each sample a pair of contacts along a line
parallel to the $x$-direction has been attached (see
Fig.~\ref{fig1}c). We use Cartesian coordinates for (001)-oriented
samples $x \parallel [1\bar{1}0]$, $y \parallel [110]$, $z
\parallel [001]$ and for (113)-oriented samples $x^\prime=x
\parallel [1\bar{1}0]$, $y^\prime \parallel [33\bar{2}]$,
$z^\prime \parallel [113]$. Right
handed ($\sigma_+$) and left handed ($\sigma_-$) circularly polarized radiation was achieved by
using a Fresnel rhomb. In order to correlate the spectral
dependence of the CPGE current to the absorption of the QWs,
optical transmission measurements were carried out  using a
Fourier transform infrared spectrometer. The current $j$ generated
by circularly polarized light in the unbiased devices was measured
via the voltage drop across a 50~$\Omega$ load resistor in a
closed circuit configuration (see Fig.~\ref{fig1}c). The voltage
in response to a laser pulse was recorded with a storage
oscilloscope.

\begin{figure}
 \centerline{\epsfxsize 70mm \epsfbox{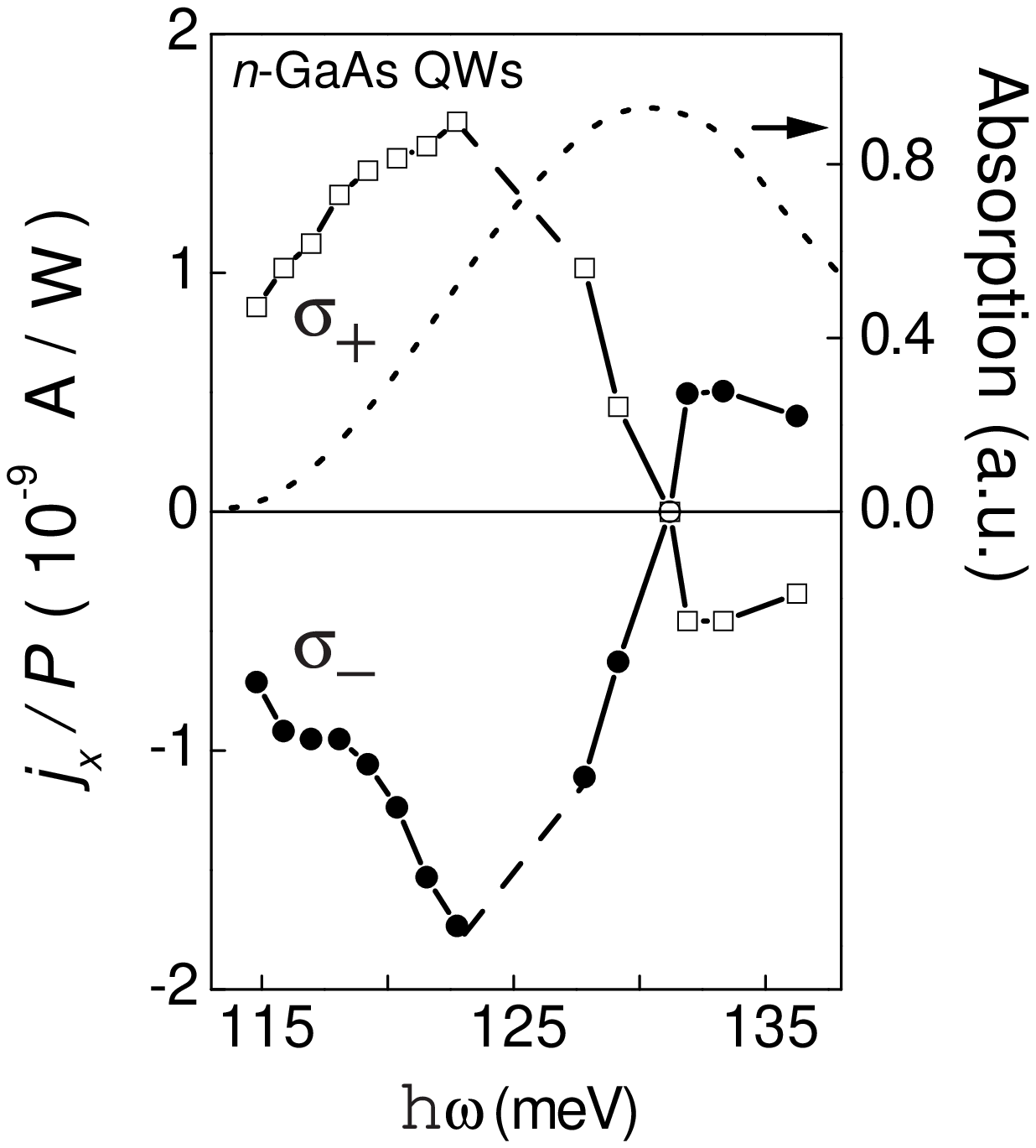  }}

\caption{ Photocurrent in QWs normalized by the light power $P$ as
a function of the photon energy $\hbar \omega$. Measurements are
presented for $n$-type (001)- grown GaAs/AlGaAs QWs of 8.2 nm
width (symmetry class C$_{2v}$) at $T=$293~K. Oblique incidence of
$\sigma_{+}$ (squares) and $\sigma_{-}$ (circles) circular
polarized radiation with an angle of incidence $\Theta_0=20^\circ$
was used. The current $j_{x}$ was measured perpendicular to the
direction of light propagation $y$. The dotted line shows the
absorption measurement using a Fourier transform infrared
spectrometer.} \label{fig2}
\end{figure}

Illuminating the unbiased QW structures with circularly polarized radiation
results in a current signal due to CPGE which
is proportional to the helicity $P_{circ}$ of the radiation. The
signal follows the temporal structure of the laser pulse and
changes sign if the circular polarization is switched from
$\sigma_+$ to $\sigma_-$. Typical signal traces are shown in
Fig.~\ref{fig1} compared to records of a linear photon
drag detector~\cite{Ganichev84p20}. In (001)- oriented samples,
belonging to the point group C$_{2v}$, the CPGE current   is only
observed under oblique incidence of radiation, as expected from
symmetry~\cite{PRL01}. For illumination along $y$-direction the
helicity dependent photocurrent flows in $x$-direction
perpendicular to the wavevector of the incident light. This is observed in experiment.
In Fig.~\ref{fig2} the photocurrent as a function of photon energy is
plotted for $\sigma_+$ and $\sigma_-$ polarized radiation together
with the absorption spectrum. The data are presented for a
(001)-grown $n$-GaAs QW of 8.2~nm width measured at room
temperature. The current for both, left and right handed circular
polarizations, changes sign at a frequency $\omega =
\omega_{inv}$. This inversion frequency $\omega_{inv}$ coincides
with the frequency of the absorption peak (see Fig.~\ref{fig2}).
The absorption peak frequency and  $\omega_{inv}$ depend on the
sample width according to the variation of the subband energy
separation. This has been verified  by measuring QWs of different
widths. Spin orientation induced CPGE and its spectral sign
inversion have also been detected  in a (113)-oriented $n$-GaAs QW
which belongs to the point group C$_s$. In this case the helicity
dependent signal is observed in $x$-direction at normal incidence
of radiation along $z^\prime$.

\section{Microscopic model}

\begin{figure}
 \centerline{\epsfxsize 140mm \epsfbox{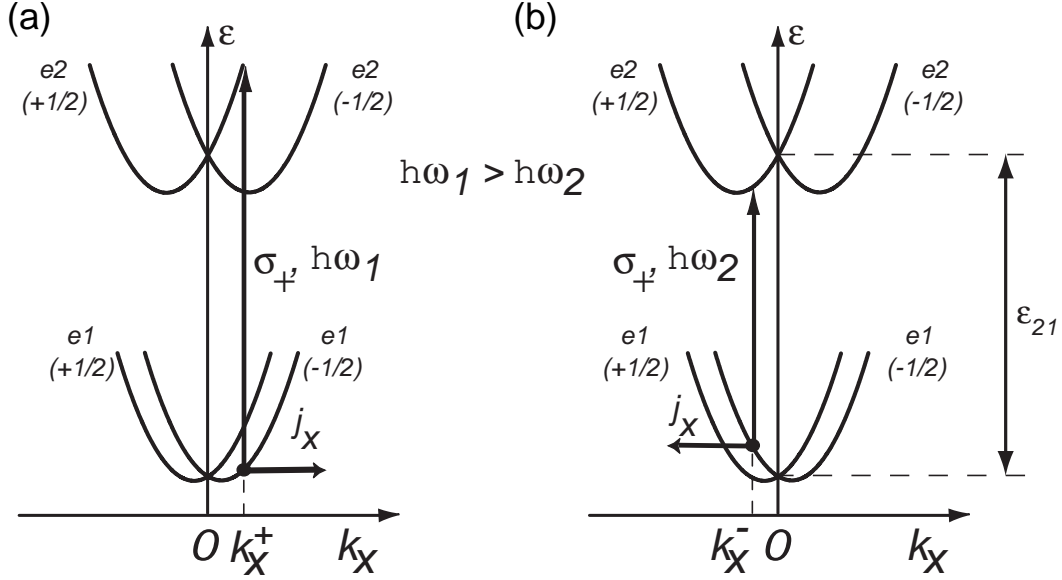  }}

\caption{Microscopic picture describing the origin of the
inversion of the photocurrent in C$_s$ point group samples. The
essential ingredient is the splitting of the conduction band due
to {\boldmath $k$}-linear terms. Right handed circularly polarized radiation, $\sigma_+$, induces
direct spin-flip transitions (vertical arrows) from $e1$ subband with  $s =
-1/2$ to $e2$ subband with $s=+1/2$. As a result an unbalanced
occupation of the $k_x$ states occurs yielding a spin polarized
photocurrent. (a) For transitions with $k_x^+$ right to the minimum of
$e1$ ($s$=-1/2) subband the current indicated by $j_x$ is
positive.  (b) At smaller $\hbar \omega$ the transition occurs at
$k_x^-$, now left to the subband minimum, and the current
reverses its sign.} \label{fig3}
\end{figure}

\begin{figure}
 \centerline{\epsfxsize 140mm \epsfbox{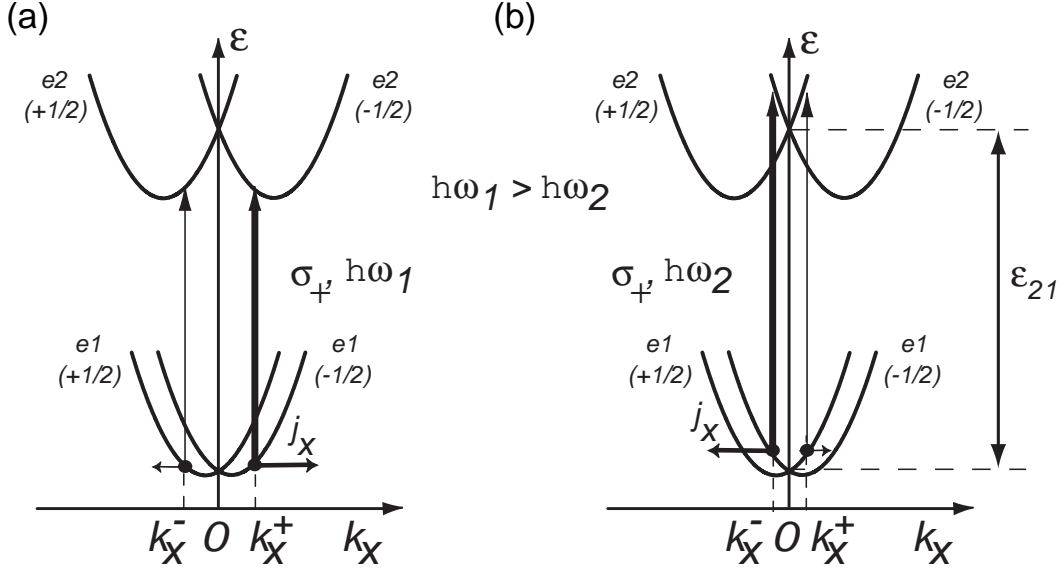  }}

\caption{Microscopic picture describing the origin of the
inversion of the photocurrent in C$_{2v}$ point group samples. (a)
Excitation with $\sigma_+$ radiation of $\hbar \omega$ less than
the energy subband separation $\varepsilon_{21}$ induces direct
spin-conserving transitions (vertical arrows) at $k_x^-$ and
$k_x^+$. The rates of these transitions  are different as
illustrated by different thickness of the  arrows (reversing the
angle of incidence mirrors the  transition rates). This leads to a
photocurrent due to an asymmetrical distribution of carriers in
{\boldmath $k$}-space if the splitting of the $e1$ and $e2$
subbands are non-equal. (b) Increasing of the photon energy shifts
more intensive transitions to the left and less intensive to the
right resulting in a current sign change.} \label{fig4}
\end{figure}

The physical origin of the effect is sketched in Fig.~\ref{fig3}
for C$_s$  symmetry and in Fig.~\ref{fig4} for C$_{2v}$ symmetry.
For both symmetries the degeneracy in {\boldmath $k$}-space is
lifted. First we consider the simplest case of C$_s$ symmetry
relevant for (113)-oriented samples. The $\sigma_{z^\prime} k_x$
contribution to the Hamiltonian, responsible for the effect under
normal incidence, splits the electron spectrum into spin
sub-levels with the spin components $s=\pm 1/2$ along the growth
direction $z^\prime$.   As a result of optical selection rules
right-handed circular polarization under normal incidence induces
direct optical transitions between the subband $e1$ with spin
$s=-1/2$ and $e2$ with spin $s=+1/2$. For monochromatic radiation
optical transitions  occur only at a fixed $k_x$ where the energy
of the incident light matches the transition energy as is
indicated by the arrow in Fig.~\ref{fig3}a. Therefore optical
transitions induce an imbalance of momentum distribution in both
subbands yielding an electric current. However, a non-equilibrium
distribution of carriers in the upper subband rapidly relaxes due
to the very effective relaxation channel of $LO$-phonons emission,
because the energy separation $\varepsilon_{21}$ between $e1$ and
$e2$ at $k_x =0$ is well above the energy of $LO$ phonons in
$n$-GaAs QWs ($\varepsilon_{LO}$= 35.4~meV).
 Therefore
the contribution of the $e2$ subband to the electric current
vanishes and the magnitude and direction of electron flow is
determined by the momentum distribution of carriers in the lowest
subband.

Fig.~\ref{fig3}a and \ref{fig3}b show what happens when, as in our
experiment,  the energy of the incident light is varied from
energies above $\varepsilon_{21}$ to values below
$\varepsilon_{21}$. Here $\varepsilon_{21}$ is the subbands'
energy separation at {\boldmath$k$}=0. At large photon energy,
$\hbar\omega > \varepsilon_{21}$, excitation occurs  at positive
$k_x$ resulting in a current $j_x$ shown by arrow in
Fig.~\ref{fig3}a. A reduced photon frequency shifts the transition
towards negative $k_x$ and reverses the direction of the current
(Fig.~\ref{fig3}b). The inversion of the current's sign occurs at
a photon frequency $\omega_{inv}$ corresponding to the transition
at the minimum of $e1$ (s=-1/2). The model suggests that the
magnitude of the spin splitting described by the Rashba and
Dresselhaus terms~\cite{Bychkov84p78,Dyakonov86p110} could easily
be derived from the energy shift  $\hbar\omega_{inv} -
\varepsilon_{21}$. However, our measurements show, that the
inhomogeneous broadening of the optical transition in real QWs is
too large to obtain this energy shift. On the other hand the model
clearly shows that without {\boldmath $k$}-linear terms in the
band structure neither an inversion nor a  current would exist.
Similar arguments hold for C$_{2v}$ symmetry (relevant for
(001)-oriented samples) under oblique incidence (see
Fig.~\ref{fig4}) although the simple selection rules are no longer
valid~\cite{Warburton96p7903}. This is pointed out in more detail
at the end of the next section.

\section{Microscopic theory}

The theory of the circular photogalvanic effect is developed by using the spin density matrix
technique ~\cite{book}.
Generally the total electric current that appears in a structure
under inter-subband excitation consists of the contributions from the
$e1$ and $e2$ subbands
\begin{equation}
{\mbox{\boldmath$j$}}={\mbox{\boldmath$j$}}^{(1)}+{\mbox{\boldmath$j$}}^{(2)} \:,
\label{eq1}
\end{equation}
which in the relaxation time approximation are given by standard
expressions
\begin{equation}
{\mbox{\boldmath$j$}}^{(\nu)}=e  \sum_{\mbox{\boldmath$k$}} \tau^{(\nu)}_p \mbox{Tr} \left[\,
\hat{\mbox{\boldmath$v$}}^{(\nu)} ({\mbox{\boldmath$k$}})\, \dot{\rho}^{(\nu)}({\mbox{\boldmath$k$}})
\right]\:. \label{eq2}
\end{equation}
Here $e$ is the electron charge, $\nu$=1,2 labels the subband $e\nu$, $\tau^{(\nu)}_p$ is the momentum
relaxation time in the subband $e\nu$, $\dot{\rho}^{(\nu)}({\mbox{\boldmath$k$}})$
is the generation of the density matrix and
$\hat{{\mbox{\boldmath$v$}}}^{(\nu)}$ is the
velocity operator in the subband given by
\begin{equation}
\hat{{\mbox{\boldmath$v$}}}^{(\nu)}=\hbar^{-1}\,\nabla_{\bm{k}}\hat{H}^{(\nu)}
\:.
\end{equation}
For the sake of simplicity we will consider a parabolic electron
spectrum for the subbands and take a Hamiltonian of the form
\begin{equation}
\hat{H}^{(\nu)}=\varepsilon^{(\nu)} + \frac{\hbar^2 { \mbox{\boldmath$k$}}^2}{2m^*} +
\hat{\mathcal{H}}^{(\nu)}_{1\bm{k}} \:,
\end{equation}
where $\varepsilon^{(\nu)}$ is the energy of size-quantization,
$\hat{\mathcal{H}}^{(\nu)}_{1\bm{k}}$ is the spin-dependent ${\mbox{\boldmath$k$}}$-linear
contribution, and $m^*$ is the effective mass equal for both
subbands.

It is convenient to write the generation matrices in the basis of
spin eigenstates $\chi_{\nu\bm{k}s}$ of the Hamiltonians
$\hat{H}^{(\nu)}$. For the case of inter-subband transitions $e1
\rightarrow e2$ the corresponding equations have the form
(see~\cite{ILP})
\begin{equation}\label{eq5}
\dot{\rho}^{(1)}_{ss^\prime}({\mbox{\boldmath$k$}})=-\frac{\pi}{\hbar}
\sum_{s_1} \mathcal{M}_{s_1,\,s^\prime}({\mbox{\boldmath$k$}})\,
\mathcal{M}^*_{s_1,\,s}({\mbox{\boldmath$k$}}) \times
\end{equation}
\[
\left[ f_{\bm{k}s} \delta(\varepsilon_{2\bm{k}s_1}-\varepsilon_{1\bm{k}s}-\hbar\omega)
+ f_{\bm{k}s^\prime}
\delta(\varepsilon_{2\bm{k}s_1}-\varepsilon_{1\bm{k}s^\prime}-\hbar\omega) \right] \:,
\]

\[
\dot{\rho}^{(2)}_{ss^\prime}(\bm{k})=\frac{\pi}{\hbar} \sum_{s_1}
f_{\bm{k}s_1} \mathcal{M}_{s,\,s_1}(\bm{k})\,
\mathcal{M}^*_{s^\prime,\,s_1}(\bm{k}) \times
\]
\[
\left[ \delta(\varepsilon_{2\bm{k}s}-\varepsilon_{1\bm{k}s_1}-\hbar\omega) +
\delta(\varepsilon_{2\bm{k}s^\prime}-\varepsilon_{1\bm{k}s_1}-\hbar\omega) \right] \:,
\]
Here $s$, $s^\prime$ and $s_1$ are the spin indices, $f_{\bm{k}s}$
is the equilibrium distribution function in the subband $e1$ (the
subband $e2$ is empty in equilibrium), $\varepsilon_{\nu \bm{k}s}$ is the
electron energy, and $\mathcal{M}_{s_1,\,s}(\bm{k})$ is the matrix
element of inter-subband optical transitions $(e1,\bm{k},s)
\rightarrow (e2,\bm{k},s_1)$. The latter is given by
$\mathcal{M}_{s,\,s_1}(\bm{k})=\chi_{2\bm{k} s}^{\dag} \,\hat{M}\,
\chi_{1\bm{k} s_1}$, where $\hat{M}$ is a $2\times2$ matrix
describing the inter-subband transitions in the basis of fixed
spin states $s_z = \pm 1/2$,
\begin{equation}\label{eq18}
\hat{M} = -\frac{eA}{cm^*}\,p_{21}\left[
\begin{array}{cc} e_z & \Lambda (e_x - i e_y) \\ - \Lambda (e_x + i
e_y) & e_z
\end{array} \right]\;,
\end{equation}
$A$ is the amplitude of the electro-magnetic wave related to light
intensity by $I=A^2\omega^2 n_{\omega} / (2 \pi c)$,
{\boldmath$e$} is the unit vector of the light polarization,
$n_{\omega}$ is the refraction index of the media, $c$ is the
light velocity, and $p_{21}$ is the momentum matrix element
between the envelope functions of size quantization
$\varphi_{1}(z)$ and $\varphi_{2}(z)$ in the subbands $e1$ and
$e2$,
\begin{equation}
p_{21} = - i \hbar \int \varphi_{2}(z) \frac{\partial}{\partial z}
\varphi_{1}(z)\: dz \:.
\end{equation}
The parameter $\Lambda$ originates from $\bm{k}\cdot\bm{p}$
admixture of valence band states to the electron wave function
and is given by
\begin{equation} \Lambda = \frac{ \varepsilon_{21} \Delta (2
\varepsilon_g + \Delta)}{2\varepsilon_g (\varepsilon_g + \Delta)(3 \varepsilon_g + 2 \Delta)}\:,
\end{equation}
where $\varepsilon_g$ is the energy of the band gap, and $\Delta$ is the
energy of spin-orbit splitting of the valence band. As one can see
from Eq.~(\ref{eq18}), the parameter $\Lambda$ determines the
absorbance for the light polarized in the interface plane.

In ideal QWs the circular photocurrent $\bm{j}$ may be obtained
from Eqs.~(\ref{eq1})-(\ref{eq5}). However, in real structures the spectral width of the
inter-subband resonance is broadened due to fluctuation of the
QW width and hence exceeds the spectral width of the
absorption spectrum of an ideal structure. The inhomogeneous broadening can be taken
into account assuming that the energy separation between subbands
$\varepsilon_{21}$ varies in the QW plane. Then by convolution of the
photocurrent $\bm{j}(\varepsilon_{21})$ with the distribution function
$F(\varepsilon_{21})$ we obtain
\begin{equation}
\bar{\bm{j}}=\int \bm{j}(\varepsilon_{21})\, F(\varepsilon_{21})\, d\varepsilon_{21} \:.
\end{equation}
The function $F(\varepsilon_{21})$ for inhomogeneous broadening may be expanded in
powers of $\varepsilon_{21} - \hbar \omega$ and by considering only the first two terms we obtain
\begin{equation}
F(\varepsilon_{21}) \approx F(\hbar\omega) + F^\prime(\hbar\omega)
(\varepsilon_{21}-\hbar\omega) \:.
\end{equation}
Taking into account the Hamiltonian
$\hat{\mathcal{H}}^{(\nu)}_{1\bm{k}}$ to be linear in $\bm{k}$,
the averaged current is finally given by
$$
\bar{\bm{j}}=e n_e \frac{\pi}{\hbar^2} \left[ \tau^{(2)}_p
F(\hbar\omega) + \left( \tau^{(1)}_p-\tau^{(2)}_p \right)
F^\prime(\hbar\omega) \bar{\varepsilon} \right] \times
$$
\begin{equation}
\mbox{Tr} \left[ \hat{M}^{\dag}
\left(\nabla_{\bm{k}}\hat{\mathcal{H}}^{(2)}_{1\bm{k}}\right)
\hat{M} - \hat{M}
\left(\nabla_{\bm{k}}\hat{\mathcal{H}}^{(1)}_{1\bm{k}}\right)
\hat{M}^{\dag} \right] \:,
\label{eq9}
\end{equation}
where $n_e$ is the 2D carrier density, and $\bar{\varepsilon}$ is the mean
value of the electron energy. For a degenerate 2D electron gas
$\bar{\varepsilon} = \varepsilon_F/2$ and for a non-degenerate gas $\bar{\varepsilon} = k_BT$,
where $\varepsilon_F$ is the Fermi energy, $k_B$ is the Boltzmann constant
and $T$ is the temperature. We note, that the distribution
function $F(\hbar\omega)$ determines the spectral behaviour
of the absorbance in the presence of an inhomogeneous broadening.

\subsection{C$_s$- symmetry and normal incidence.}

In $(113)$-grown QW structures of C$_s$ symmetry the CPGE occurs
under normal incidence of the radiation.  In this case the
$\bm{k}$--linear contribution to the Hamiltonian responsible for
the effect is given by $\beta_{z^\prime x}^{(\nu)}
\sigma_{z^\prime} k_x$. Here $\beta_{z^\prime x}^{(\nu)}$ are the
coefficients being different for the $e1$- and $e2$ subbands. The
$\bm{k}$-linear term splits the electron spectrum into spin
sub-levels with the spin components $s = \pm 1/2$ along the growth
direction $z^\prime$ (see Fig.~\ref{fig3}). Thus the electron
parabolic dispersion in the subbands $e1$ and $e2$ has the form
\begin{equation}
\varepsilon_{\nu, \bm{k}, \pm 1/2} = \varepsilon^{(\nu)} + \frac{\hbar^2 \left(k_{x}^2
+ k_{y^\prime}^2\right)}{2m^*} \pm \beta_{z^\prime x}^{(\nu)}
k_{x}\:.
\label{eq10}
\end{equation}
For direct inter-subband transitions under normal incidence
selection rules allow only the spin-flip transitions, $(e1, -1/2)
\rightarrow (e2, 1/2)$ for $\sigma_+$ photons and $(e1, 1/2)
\rightarrow (e2, - 1/2)$ for $\sigma_-$
photons~\cite{Warburton96p7903}.
Due to these selection rules together with energy and momentum conservation
laws the optical inter-subband transition under, for example, $\sigma_+$
photoexcitation is only allowed for the fixed wavevector $k_x$ given by
\begin{equation} \label{eq11}
k_{x} = \frac{\hbar \omega - \varepsilon_{21}}{\beta_{z^\prime x}^{(2)} +
\beta_{z^\prime x}^{(1)}}\:.
\end{equation}
Velocities of electrons in the $e2$ subband and of 'holes' in the $e1$ subband generated by this transition
are given  by
\begin{equation} \label{eq12}
v^{(1)}_{x} = \hbar k_{x} / m^* - \beta_{z^\prime x}^{(1)} /\hbar
\:,\:\:\:\:\:\:\: v^{(2)}_{x} = \hbar k_{x} / m^* + \beta_{z^\prime x}^{(2)}
/\hbar \:,
\end{equation}
This unbalanced distribution of carriers in {\boldmath$k$}-space induces an electric current
\begin{equation} \label{eq13}
j_{x} = j^{(1)}_x + j^{(2)}_x =-e \frac{\eta_{\parallel} I}{\hbar \omega} \left(
v^{(1)}_{x} \tau_p^{(1)} - v^{(2)}_{x} \tau_p^{(2)}
\right) P_{circ} \:,
\end{equation}
where $I$ is the light intensity and $\eta_{\parallel}$ is the
absorbance or the fraction of the energy flux absorbed in the QW
due to the inter-subband transitions  under normal incidence.
 Note that the magnitude of the photocurrent $j^{(2)}_{x}$,
 corresponding to the second term in the bracket of Eq.~(\ref{eq13}),
 and stems from photoelectrons in the $e2$
subband is smaller than $\left|j^{(1)}_{x}\right|$ because $\tau^{(2)}_p <
\tau^{(1)}_p$ as described above. The resonant inversion of the
circular photocurrent is clearly seen from Eqs.~({\ref{eq11}) -
(\ref{eq13}) because $\eta_{\parallel}$ is positive and
$v^{(\nu)}_{x}$ changes its sign at a particular frequency.

For a degenerate 2D electron gas at low temperature we find that the dependence of
the absorbance $\eta_{\parallel}$
on $\hbar \omega$ and $\beta^{(\nu)}_{z^\prime x}$ is given by
\begin{equation} \label{eq14}
\frac{\eta_{\parallel}}{\hbar \omega} \propto
\frac{1}{\left|\beta_{z^\prime x}^{(2)} + \beta_{z^\prime x}^{(1)}\right|}
\left[ \tilde{\varepsilon}_F - \frac{\hbar^2}{2 m^*} \left( \frac{\hbar
\omega - \varepsilon_{21}}{\beta_{z^\prime x}^{(2)} + \beta_{z^\prime x}^{(1)}}
\right)^2\right]^{1/2}\:,
\end{equation}
where $\tilde{\varepsilon}_F = \varepsilon_F - m^*[\beta_{z^\prime x}^{(1)}/ (\sqrt{2} \hbar)]^2$.

Taking into account the inhomogeneous broadening we finally obtain
for the averaged circular photocurrent
\begin{equation}\label{eq15}
\bar{j}_x = \frac{e}{\hbar} (\beta_{z^\prime x}^{(2)} +
\beta_{z^\prime x}^{(1)}) \left[ \tau_p^{(2)}\:
\bar{\eta}_{\parallel}(\hbar \omega) + (\tau_p^{(1)} -
\tau_p^{(2)})\: \bar{\varepsilon}\: \frac{d \:\bar{\eta}_{\parallel}(\hbar
\omega)}{d\: \hbar \omega} \right] \frac{IP_{circ}}{\hbar \omega}  \:,
\end{equation}
where $\bar{\eta}_{\parallel} \propto F(\hbar \omega)$ is the calculated
absorbance  neglecting $\bm{k}$-linear terms but
taking into account the inhomogeneous broadening.

\subsection*{C$_{2v}$ symmetry and oblique incidence.}

In the case of C$_{2v}$ point symmetry which is relevant for (001)-oriented
QWs the current flows only at oblique incidence and  is caused by $\bm{k}$-linear
contributions to the electron effective Hamiltonian given by
\begin{equation} \label{eq16}
{\cal{H}}^{(\nu)}_{1{\bm k}} = \beta^{(\nu)}_{xy} \sigma_x k_y +
\beta^{(\nu)}_{yx} \sigma_y k_x\:.
\end{equation}
The coefficients $\beta^{(\nu)}_{xy}$ and $\beta^{(\nu)}_{yx}$ are
related to the bulk-inversion asymmetry (BIA) or Dresselhaus
term~\cite{Dyakonov86p110} and structure-inversion asymmetry (SIA)
or Rashba term~\cite{Bychkov84p78} by
\begin{equation} \label{eq17}
\beta^{(\nu)}_{xy} = \beta^{(\nu)}_{BIA} +
\beta^{(\nu)}_{SIA}\:,\:\:\:\:\:\:\:\: \beta^{(\nu)}_{yx} = \beta^{(\nu)}_{BIA}
- \beta^{(\nu)}_{SIA}\:.
\end{equation}

The circular photocurrent due to inter-subband transition in
(001)-grown QWs in the presence of inhomogeneous broadening can be
calculated following Eq.~(\ref{eq9}) yielding
\begin{equation}\label{eq21}
j_x =  - \Lambda \frac{e}{\hbar} \left(\beta^{(2)}_{yx}-\beta^{(1)}_{yx}\right) \left[ \tau_p^{(2)}\:
\eta_{\perp}(\hbar \omega) + (\tau_p^{(1)} - \tau_p^{(2)})\:
\bar{\varepsilon}\: \frac{d \:\eta_{\perp}(\hbar \omega)}{d\: \hbar \omega}
\right] \frac{I P_{circ}}{\hbar \omega}\:
\hat{e}_y
\end{equation}
where $\hat{\bm{e}}$ is the unit vector directed along the light
propagation and $\eta_{\perp}$ is the absorbance for the
polarization perpendicular to the QW plane. The current in $y$-direction can be
obtained by interchanging the indices $x$ and $y$ in Eq.~(\ref{eq21}).

The origin of the spin orientation induced CPGE caused by direct
inter-subband transitions in C$_{2v}$-symmetry systems is
illustrated in Fig.~\ref{fig4} for $\sigma_+$ radiation. In
C$_{2v}$-symmetry the $\sigma_{y} k_x$  contribution to the
Hamiltonian splits the subbands in $k_x$ direction in two spin
branches with $s=\pm1/2$ oriented along $y$ (see Fig.~\ref{fig4}).
Due to selection rules the absorption of circularly polarized
radiation is spin-conserving~\cite{book}. The asymmetric
distribution of photo-excited electrons resulting in a current is
caused by these  spin-conserving but spin-dependent transitions
(see Eq.~(\ref{eq21})). This is in contrast to spin-flip processes
occurring in  (113)-grown QWs described above. It turns out that
under oblique excitation by circularly polarized light the rates
of inter-subband transitions are different for electrons with the
spin oriented co-parallel and antiparallel to the in-plane
direction of light propagation~\cite{IT}. The difference  is
proportional to the product $|M_{\parallel} M_{\perp}|$, where
$M_{\parallel}$ and $M_{\perp}$ are the absorption matrix elements
for in-plane and normal light polarization. This is depicted in
Fig.~\ref{fig4} by vertical arrows of different thickness. In
systems with $\bm{k}$-linear spin splitting such processes lead to
an asymmetrical distribution of carriers in $\bm{k}$-space, i.e.
to an electrical current. Similar to C$_s$ symmetry  the variation
of the photon energy leads to the inversion of the current
direction (see Fig.~\ref{fig4}a and~\ref{fig4}b). Since the
circular photogalvanic effect in QW structures of C$_{2v}$
symmetry is caused by spin-dependent {\it spin-conserving} optical
transitions, the photocurrent described by Eq.~(\ref{eq21}) in
contrast to Eq.~(\ref{eq15}) is proportional to the difference of
subband spin splittings.

\section{Conclusions}

Our experiments show that direct  inter-subband transitions in
$n$-type GaAs QWs result in a spin orientation induced CPGE. The
central observation is a change of the sign of the
 CPGE current close to resonance of
inter-subband transitions. The theoretical results give a detailed
description of all  features observed in experiment. The sign
inversion  follows for C$_s$ symmetry from Eq.~(\ref{eq15}) and
for C$_{2v}$ symmetry from Eq.~(\ref{eq21}). If $\tau^{(2)}_p \ll
\tau^{(1)}_p$ the first term on the right hand side in square
brackets of both equations is vanishingly small compared to the
second one. Therefore the photocurrent is proportional to the
derivative of the absorbance and is zero at the frequency of the
absorption peak as observed in experiment. Thus the assumption
that the momentum relaxation in the upper subband is much faster
than in the lower subband is satisfied.

Comparing theory with experiment we also note that for C$_{2v}$ symmetry the spin splitting in
{\boldmath$k$}-space is different for the $e1$ and $e2$ subbands.
Eq.~(\ref{eq21}) shows that for equal spin splitting, $\beta^{(2)}_{yx} = \beta^{(1)}_{yx} $, the
current vanishes. For C$_s$ symmetry in contrast, the
spin-orientation induced CPGE current is proportional to the sum
of the subband spin splittings and therefore exist for $\beta^{(2)}_{yx} = \beta^{(1)}_{yx} $.
The observation of the CPGE for direct inter-subband
transitions allows to extend the method of
spin-sensitive bleaching, previously demonstrated for $p$-type
QWs~\cite{PRL02}, and therefore to investigate electron spin relaxation times
in $n$-type QWs for monopolar spin
orientation~\cite{Nature02,PRL02}.

We acknowledge financial support from the DFG, the RFBR and
the programs of the RAS.

\end{document}